# *Chandra* X-ray Observations of the Neutron Star Merger GW170817: Thermal X-Ray Emission From a Kilonova Remnant?

by Samar Safi-Harb[1,*], Neil Doerksen[1], Adam Rogers[1,2] and Chris L. Fryer[3]

[1] Dept of Physics & Astronomy, University of Manitoba, Winnipeg, MB R3T 2N2
[2] Dept of Physics & Astronomy, Brandon University, Brandon, MB, R7A 6A9
[3] CCS Division, Los Alamos National Laboratory, Los Alamos, NM 87545
*Corresponding author: samar.safi-harb@umanitoba.ca

## Abstract

The recent discovery of the neutron star merger and multi-messenger event GW170817 (also known as kilonova GRB170817A) provides an unprecedented laboratory in which to study these mysterious objects, as well as an opportunity to test cutting-edge theories of gravity in the strong field regime. Before this event, such tests of our understanding of the nature of gravity were not possible. In this study, we analyze the X-ray observations of GW170817 obtained with NASA's *Chandra X-ray Observatory* following the 17 August 2017 detection of the event by the Laser Interferometer Gravitational Wave Observatory (LIGO). Motivated by understanding the emission mechanism for X-ray light and the outcome of the merger event, we fit the *Chandra* data with both non-thermal (as done previously in the literature) and thermal models. We specifically explore thermal plasma models that would be expected from a kilonova remnant (KNR). We reproduced the non-thermal results which were recently published by Nynka et al. (2018). We also find that thermal bremsstrahlung emission from hot plasma can account for the X-ray emission from this source. Furthermore, we consider models allowing for an intrinsic absorption from the merger event, yielding a softer power-law model photon index than previously published, or a lower plasma temperature. We also report on evidence for line emission, or excess above the continuum model, near 1.3 keV and 2.2 keV which sheds new light on the interpretation of the KNR and its nucleosynthesis products. We discuss the feasibility for the KNR as the origin for thermal X-ray emission at this stage of the kilonova evolution.

## Introduction

The lives of massive stars more than eight times the mass of the Sun end violently in supernova (SN) explosions. The outer layers of these stars are scattered through interstellar space, forming complex supernova remnants (SNRs) which disperse the heavy elements (referred to as ejecta) created within the star during its evolution or via the explosion itself. Meanwhile, a portion of the stellar core survives compressed to a cinder of nuclear density. Under such intense pressures and densities, the core is converted into a neutron-rich material called neutron degenerate matter. This state, unique in the universe, provides a name for these enigmatic objects – neutron stars, which manifest themselves as a "zoo" when observed at different wavelengths (Hewish et al. 1968, Shapiro & Teukolsky 1983, Haensel, Potekhin & Yakovlev 2007, Harding 2013, Safi-Harb 2017).

Neutron stars exhibit a plethora of interesting and mysterious behavior. They emit highly collimated jets of radiation in both the X-ray and radio portions of the electromagnetic spectrum. They have some of the strongest magnetic fields in nature, up to and beyond the quantum electrodynamics limit (Duncan & Thompson 1992, Kouveliotou et al. 2003). Their association with SNRs allows us to probe some of their most fundamental properties such as magnetic field evolution (e.g., Rogers & Safi-Harb 2016). Neutron stars in binary systems can accrete matter from a main sequence companion, driving their rotation rates to millisecond periods (Backer et al. 1982). In addition, due to their high density, the effects of gravitational light bending, time dilation and redshift are substantial around these objects (Pechenick, Ftaclas & Cohen 1983, Beloborodov 2002, Rogers 2015). Given the diversity of interesting phenomena as-sociated with them, there is no doubt that neutron stars are some of the most fascinating and exotic objects in the Universe.

Isolated neutron stars occupy the frontiers of our understanding of nature. However, the physical description of two such objects colliding is even more extreme. Since neutron stars are so compact, they provide ideal conditions for studying gravitational waves when they are found in binary systems. The first double-pulsar system discovered was observed to lose energy in accordance with the emission of gravitational waves predicted in 1916 by Einstein's theory of general relativity. This is the Hulse and Taylor binary (Hulse & Taylor 1974) which earned its discoverers the Nobel Prize in Physics in 1993. However, until the recent success of the Laser Interferometer Gravitational Wave Observatory (LIGO), gravitational waves had not been directly observed.

The direct detection of gravitational waves remained elusive for a century until the LIGO team detected them in 2015 from black-hole mergers, earning Weiss, Barish and Thorne the 2017 Nobel Prize in Physics. In addition to black-hole binaries, GW170817 (a neutron star-neutron star merger) is one of the most exciting discoveries in the field of astronomy today. GW170817 is the first event of any kind in which both gravitational and electromagnetic waves were detected from the same event. It is also the first merger event to emit gravitational waves that involved objects other than black holes. The discovery of gravitational waves and light across the electromagnetic spectrum has led to the birth of a new era in "multi-wavelength, multi-messenger astrophysics" (Abbott et al. 2017a, b and references therein).

Following the LIGO discovery, the gamma-ray satellite Fermi detected the source seconds after the gravitational wave event, showing that GW170817 shares common properties to the short-duration gamma-ray bursts (Troja et al. 2017). The event is also referred to as GRB170817A. Optical and near-infrared telescopes located the source (Coulter et al. 2017, Shappee et al. 2017, Drout et al. 2017) within the host galaxy NGC 4993 (at a redshift z=0.0098) and a distance of about 41 Mpc (Hjorth et al. 2017). In X-rays, the source was first detected ~9 days post-merger (Troja et al. 2017, Margutti et al. 2017) and subsequently followed up at ~15-16 days post-merger (Haggard et al. 2017) using NASA's *Chandra* X-ray Observatory. Thanks to its superb imaging resolution, the gravitational wave event was also localized in X-rays by *Chandra*. The source continued to be monitored in X-rays and radio but it is now believed to be fading in the X-ray band (Nynka et al. 2018, Ruan et al. 2018, Haggard et al. 2018, Burnichon et al. 2018).

Despite the intense multi-wavelength campaign observing this event, the exact details of the aftermath of this neutron star-neutron star collision remain unclear. The production of a kilonova was observed (Arcavi et al. 2017), though the overall flux of this event was dimmer than expected (Troja et al. 2017). Interpretation of the observations include a synchrotron afterglow or the launching of an off-axis jet of material from within the cloud of debris produced during the collision. Over a year later it is uncertain what remnant the neutron star merger has produced (e.g., Nynka et al. 2018, Pooley et al. 2018).

Motivated by understanding the product of this double neutron star merger (i.e. whether it made a heavy neutron star or a light black hole; see Pooley et al. 2018) and searching for nucleosynthesis products in X-rays, we have revisited the archival *Chandra* X-ray observations of GW170817. In our study, we model the data with the commonly assumed non-thermal, power-law models, as well as explore models of thermal plasma origin. We are able to reproduce previous published work on the non-thermal interpretation (Nynka et al. 2018). We compare our themal models with those published in Ruan et al. (2018), and discuss thermal X-ray emission from a kilonova remnant candidate (KNR) that opens a new window for interpreting the aftermath of this merger.

## Observations and Methodology

The observations used to analyze GW170817 were taken with the Advanced CCD Imaging Spectrometer (ACIS) on board *Chandra*. We analyzed the archival observations summarized in Table 1. In our work, we highlight the January 2018 observations (ObsID: 20936/37/38/39/45, PI Wilkes) which were the brightest, allowing us to discover evidence for the presence of thermal X-ray emission.

The level two event files of the observations were processed using the *Chandra* Interactive Analaysis Software (CIAO) v7.6. The RGB image presented in **Fig. 1** was generated using the software DS9 v4.10. The energy band is 0.5-7 keV and the colours are chosen to reflect the low (red=0.5-1.2 keV), medium (green=1.2-2 keV) and high (blue=2-7 keV) energy X-rays.

For processing the spectral data, CIAO was used to merge spectra from the same epoch (Table 1) to increase statistical results. The source spectrum was chosen from a region centered at the peak emission from GW170817 (RA=$13^h$ $09^m$ $48.077^s$; Decl= $-23^o$ 22' 53.459", J2000) with an extraction radius of 1.97" (which corresponds to 90% encircled energy fraction near the *Chandra* on-axis position), as done in Haggard et al. (2017) and Nynka et al. (2018) (see also Margutti et al. 2017 and Troja et al. 2017). XSPEC (Arnaud 1996) v12.10.0c was used to model the data. The background was subtracted from a source-free region close to the GW event from the same CCD chip. The redshift (z) was fixed at 0.0098 (Hjorth 2017) and the column density ($N_H$) was fixed at $7.5 \times 10^{20}$ $cm^{-2}$ to compare our results to the most recent results by Nynka et al. (2018). The analyzed energy range was 0.3-8 keV for flux and the luminosity is quoted in the 0.5-10 keV band. Each spectrum was modeled using a power-law (for the non-thermal interpretation) and the Bremsstrahlung model (for the thermal interpretation).

Using the above values as input, we searched for the lowest reduced $\chi^2$ value by trying a range of minimum counts per bin when grouping the spectrum, depending on the statistics of the observation. Cash statistics were also used for the low-count data, particularly the early observations (ObsID: 18955, 19294). Flux, luminosity, and count rates were recorded for the non-thermal and thermal models, with the power-law photon index, $\Gamma_x$, and plasma temperature (kT in keV) summarized for the power-law and Bremsstrahlung models, respectively. Our results are summarized in **Table 1**.

## Results

**Fig. 2** shows the best fit power-law and Bremsstrahlung model fits to the January 2018 merged spectra (corresponding to near-peak flux). The corresponding power-law photon index is 1.67 (1.39-1.95) and the thermal Bremsstrahlung temperature is 6.6 (3.8-17) keV, with the ranges quoted at the $2\sigma$ confidence level. These results are obtained when fixing the column density $N_H$ to 7.5 x $10^{20}$ cm$^{-2}$. When allowing $N_H$ to vary, we find a higher column density, a softer power law index, or a lower temperature for the thermal Bremsstrahlung models. In particular, for the January 2018 epoch observations, our power-law model fit yields $N_H$ = 8.9 (2.3-16) x $10^{21}$ cm$^{-2}$ (for *Wilm* abundances in XSPEC, Wilms et al. 2000) and a photon index of $\Gamma$ = 2.3 (1.7-3.0). For the thermal Bremsstrahlung model, we find $N_H$ = 5 (0.22-10) x $10^{21}$ cm$^{-2}$ and a temperature of 3.7 (2.2-9.7) keV. These results suggest an additional intrinsic component from the merger absorbing the softer X-rays.

By comparing our power-law model fit results with those of Nynka et al. (2018), where the column density was fixed to 7.5 x $10^{20}$ cm$^{-2}$, we find that our values are consistent with theirs within error. Our thermal Bremsstrahlung model fits also yield acceptable fits, and in some cases, are statistically preferred over the power-law models. This suggests that the post-merger event can be described by hot plasma (kT~6.6 keV for the January 2018 observations), which was not considered previously in the literature. These temperatures would be lower (kT~3.7 keV) when allowing the column density to fit freely, as mentioned above. We note that Ruan et al. (2018) tested thermal blackbody models and found kT = 0.63 +/- 0.09 keV; however they disfavored this model for explaining both the X-ray and radio emission.

**Fig. 3** shows the evolution of the flux in both models (power-law and Bremsstrahlung) and highlights the comparison between our and the Nynka et al.'s results, confirming the consistency in our power-law model fits.

By further examining the brightest, January 2018 observations fitted with a power-law model, we note an excess of photons at energies around 1.3 keV and 2.2 keV. This is shown in **Fig.4** which illustrates the deviation from a pure power-law model. For this spectrum, the grouping was done by a minimum of 10 counts per bin and the column density was fixed to 7.5 x $10^{20}$ cm$^{-2}$. Adding two Gaussian lines near these energies improves the fit (from a reduced $\chi^2$ of 1.177 (12 degrees of freedom) to 0.738 (8 degrees of freedom). However, the line centroids and equivalent widths are poorly constrained despite this being the brightest observation with the most counts (~150 counts in the 0.3-8 keV energy range). The limited number of counts prohibits us from constraining the model parameters. When fixing the widths of the lines, we find that the centroid of the line near 2.3 keV can be constrained to the 2.15-2.34 keV range ($2\sigma$ confidence).

Our result is interesting but is severely limited by statistics. This is consistent with the conclusion reached by Ruan et al. (2018). To further verify our results, we tried the following: (a) subtract another background (annulus around the source), (b) examine the individual spectra (as opposed to the merged spectra), and (c) allow the column density to fit freely. While the latter test led (as expected) to a softer photon index of ~2.2 (1.7-3.0) for $N_H$ = 7 (6.6-15) x $10^{21}$ cm$^{-2}$, we find the same trend for the excess emission near 1.3-1.5 keV and between 2-3 keV, confirming the deviation from the pure power-law model. For the individual spectra, as expected, the (even) smaller number of counts prohibits us from any significant detection.

## Discussion

The above-mentioned evidence for line detection as well as the thermal Bremsstrahlung model fits suggest a thermal origin or an additional thermal component for the X-ray emission, which would be at odds with the assumption that the X-ray and radio population arise from the same population of synchrotron emitting plasma that supports an off-axis structured jet (e.g., Alexander et al. 2018). When allowing the column density to fit freely, we find a much higher value than that assumed in the literature, suggesting internal absorption by the merger event (see also Pooley et al. 2018). Furthermore, in this model, the power-law photon index is $\Gamma \sim 2.2$, i.e. steeper than the value of $\Gamma \sim 1.6$ at the fixed column density. The line energies, or at least the excess emission detected in the 1-2 keV and 2-3 keV bands when fitting with a continuum model, hint at the presence of shock-heated ejecta from the kilonova remnant (KNR).

Next we discuss the feasibility for the KNR as the origin for thermal X-ray emission at this stage of the kilonova evolution. In Supernovae (SNe), it takes 10-100 y before the supernova light-curve evolves into a standard remnant with a reverse shock and high-temperature ions. However, the evolution of the KNR is much faster than a normal SN. Typical kilonova ejecta masses are 1000 times less than SNe and their expansion velocities are 10 times higher (Wollaeger et al. 2018). A key indicator of the remnant evolution is the timescale to sweep up the equivalent ejecta mass. The swept-up mass, $M_{sw}$:

$$M_{sw} = 4\pi/3 \; \rho_{CSM} \, (v_{ejecta} t)^3 \; ;$$

where $t$ is the time, $\rho_{CSM}$ is the circumstellar medium density (this medium is typically 10-10,000 times less dense for KNe than SNe; e.g., Fong et al. 2010) and $v_{ejecta}$ is the ejecta velocity (the dynamical ejecta velocity is 0.2–0.3 times the speed of light, 10 times faster than SNe). The time to sweep-up an equivalent mass is:

$$t_{equiv} = (3M_{ej}/4\pi\rho_{CSM}v^3)^{1/3} \; ;$$

10-100 times shorter for KNe than SNe. The KNR will produce a reverse shock 10-100 times sooner than a SNR. This simple estimate suggests that a 100-day old KNR would have the properties of a 10 yr-old SNR.

To understand this better, let's look at one of the key properties in remnant emission, the decoupling of ions and electrons. When ions and electrons decouple (for instance, when the coupling timescale is a fraction of the remnant age), the ions can get sufficiently hot to drive X-ray emission. The coupling timescale is inversely proportional to the density of the shock. Fitting formulae to plasma-kinetics calculations can be used to estimate this coupling time. Using the formulae in Gericke et al. (2002), we find that the ion-electron equilibration time for a 10 eV ion is 5 days, 50 days and 500 days roughly for densities of $10^{-20}$, $10^{-21}$, $10^{-22}$ g cm$^{-3}$, respectively. For our typical SNRs, the density at the shock front drops below $10^{-20}$ g cm$^{-3}$ only at 10-30 yr. The density of the shock front for a kilonova can be less than $10^{-22}$ g cm$^{-3}$ at 10 days. This is illustrated in **Fig. 5**. As the remnant sweeps up mass and decelerates, the kilonova will produce both thermal and synchrotron emission at 100 days that is comparable with an SNR more than 100 yrs old.

We conclude that the thermal emission could come from the KNR. However, a more detailed analysis and simulations of the evolution of a KNR will be left to a future, more-in-depth statistical study that also takes into account all observations acquired to-date on this fascinating source.

## Prospects for Future Studies

Upgraded LIGO will come online in 2019 and promises to discover more such events. As these events become more routinely observed in the gravitational wave field, questions regarding the nature of the merger remnant may be answered. For now, GW170817 remains the only known neutron star merger that happens to be a nearby source. It is also the first event that confirmed the connection between neutron star mergers and short-duration gamma-ray bursts. X-ray observations have shed new light on the nature of this merger and its evolution. The evidence for thermal X-ray emission reported here stresses the importance of, not just the need to follow-up these sources as they evolve, but also the need for deep imaging and spectroscopic X-ray observations to constrain any thermal emission that would arise from shock-heated ejecta or the interaction with the surrounding medium.

Despite the ambiguity of the end-state of the collision, GW170817 heralds the dawn of a new era in multi-messenger astronomy with applications in many fields, including cosmology (e.g., Wiggins et al. 2018), and holds special promise for neutron star studies using X-ray observations with excellent spatial resolution such as *Chandra* and in the future with missions like the proposed *AXIS* (Mushotzky et al. 2018) and *Lynx* (the Lynx Team 2018).

## Acknowledgments

We acknowledge support from the Natural Sciences and Engineering Research Council of Canada (NSERC), and from the University of Manitoba's Faculty of Science Undergraduate Summer Research Award (USRA, N. Doerksen). We thank Daryl Haggard for discussions and valuable comments on the manuscript. Thanks to Ben Airey for assistance with code development for Chandra data analysis. This research made use of NASA's Astrophysics Data System and the High-Energy Astrophysics Science Archival Research Center maintained at NASA's Goddard Space Flight Center.

| ObsID | Exposure [ks] | Post-Merge[a] [days] | Count Rate [$10^{-4}$ cts·s$^{-1}$] | +/- | Flux[b] [$10^{-14}$ erg·s$^{-1}$·cm$^{-2}$] | +/- | plasma temp. [keV] | +/- | Luminosity[c] [$10^{38}$ erg·s$^{-1}$] | +/- | Red. $\chi^2$ | cstat | D.O.F. |
|---|---|---|---|---|---|---|---|---|---|---|---|---|---|
| 19294 | 49.41 | 9.2 | 3.46 | 0 | ** | ** | ** | ** | ** | ** | ** | | ** |
| 20728 | 46.69 | 15.4 | 2.81 | 0 | 0.29 | -0.22 | 1.4 | 0 | 6.48 | -5.18 | 2.28 | 2.86 | 1 |
| 20860 20861 | 74.09 24.74 | 109.2 | 14.77 | 1.2 -1.2 | 1.85 | 0.31 -1.01 | 6.79 | 13.48 -2.99 | 41.89 | 4.51 -17.24 | 1.02 | 6.09 | 6 |
| 20936 20938 20937 20939 20945 | 31.75 15.86 20.77 22.25 14.22 | 159.7 | 14.12 | 1.16 -1.16 | 1.85 | 0.68 -0.29 | 6.52 | 15.53 -2.97 | 41.68 | 4.08 -18.55 | 1 | 7.14 | 7 |
| 21080 21090 | 50.78 46 | 260 | 8.16 | 0.92 -0.92 | 1.05 | 0.11 -0.88 | 6.39 | 30.94 -3.26 | 23.71 | 4.85 -6.89 | 0.98 | 3.77 | 4 |

| ObsID | Exposure [ks] | Post-Merge[a] [days] | Count Rate [$10^{-4}$ cts·s$^{-1}$] | +/- | Flux[b] [$10^{-14}$ erg·s$^{-1}$·cm$^{-2}$] | +/- | Photon index Γ | +/- | Luminosity[c] [$10^{38}$ erg·s$^{-1}$] | +/- | Red. $\chi^2$ | cstat | D.O.F. |
|---|---|---|---|---|---|---|---|---|---|---|---|---|---|
| 19294 | 49.41 | 9.2 | 2.57 | 0 | 0.51 | -0.01 | 1.75 | 0.29 | 11.48 | 4.11 | 0.18 | 0.18 | 1 |
| 20728 | 46.69 | 15.4 | 2.81 | 0 | 0.4 | 0.88 | 2.25 | -0.07 | 9.05 | 12.31 | 2.55 | 3.36 | 1 |
| 20860 20861 | 74.09 24.74 | 109.2 | 13.66 | 1.18 -1.18 | 2.29 | 0.56 -0.37 | 1.62 | 0.3 -0.3 | 51.63 | 14.65 -10.77 | 1.24 | 7.6 | 6 |
| 20936 20938 20937 20939 20945 | 31.75 15.86 20.77 22.25 14.22 | 159.7 | 14.69 | 1.18 -1.18 | 2.35 | 0.46 -0.32 | 1.58 | 0.25 -0.25 | 53.13 | 8.29 -10.09 | 1.15 | 8.16 | 7 |
| 21080 21090 | 50.78 46 | 260 | 8.16 | 0.92 -0.92 | 1.27 | 0.37 -0.31 | 1.69 | 0.37 -0.37 | 28.68 | 5.23 -8.34 | 0.79 | 3.29 | 4 |

**Table 1.** Power law (Top) and Thermal Bremsstrahlung (Bottom) model fits for the post merger GW170817 event. The column density was fixed to 7.5 x $10^{20}$ cm$^{-2}$ and the redshift to 0.0098.
[a] Average post-merger date as in Nynka et al. (2018)
[b] Absorbed flux in the 0.3-8 keV band
[c] Luminosity in the 0.3-10 keV band

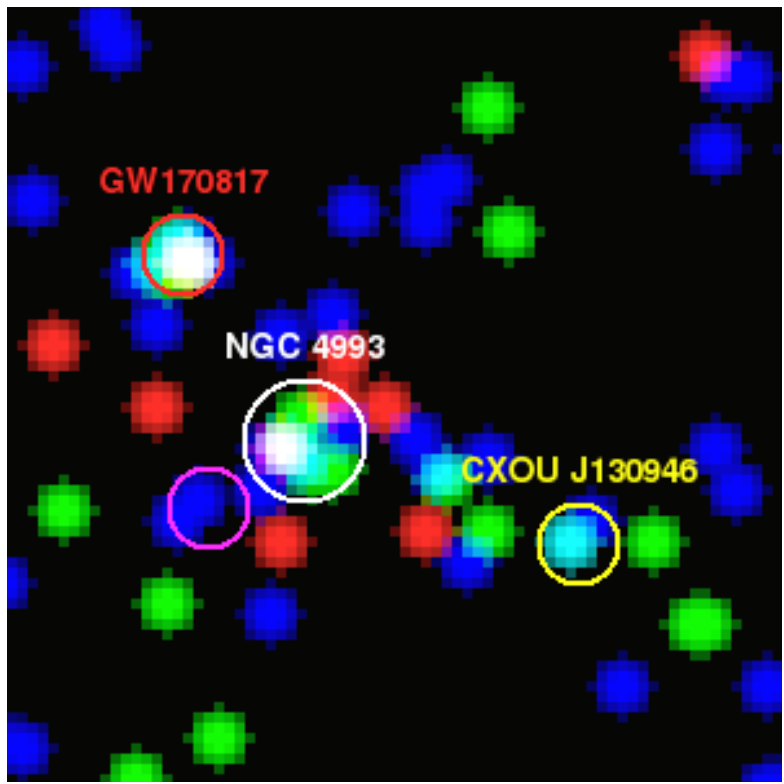

**Fig. 1:** RGB image of GW170817 at ~160 days post-merger (ObsID: 20945). Red=0.5-1.2 keV, Green=1.2-2 keV, Blue=2-7 keV. The unlabeled, magenta region is the transient source CXOU J130948; CXOU J13046 is also a transient (Haggard et al. 2017).

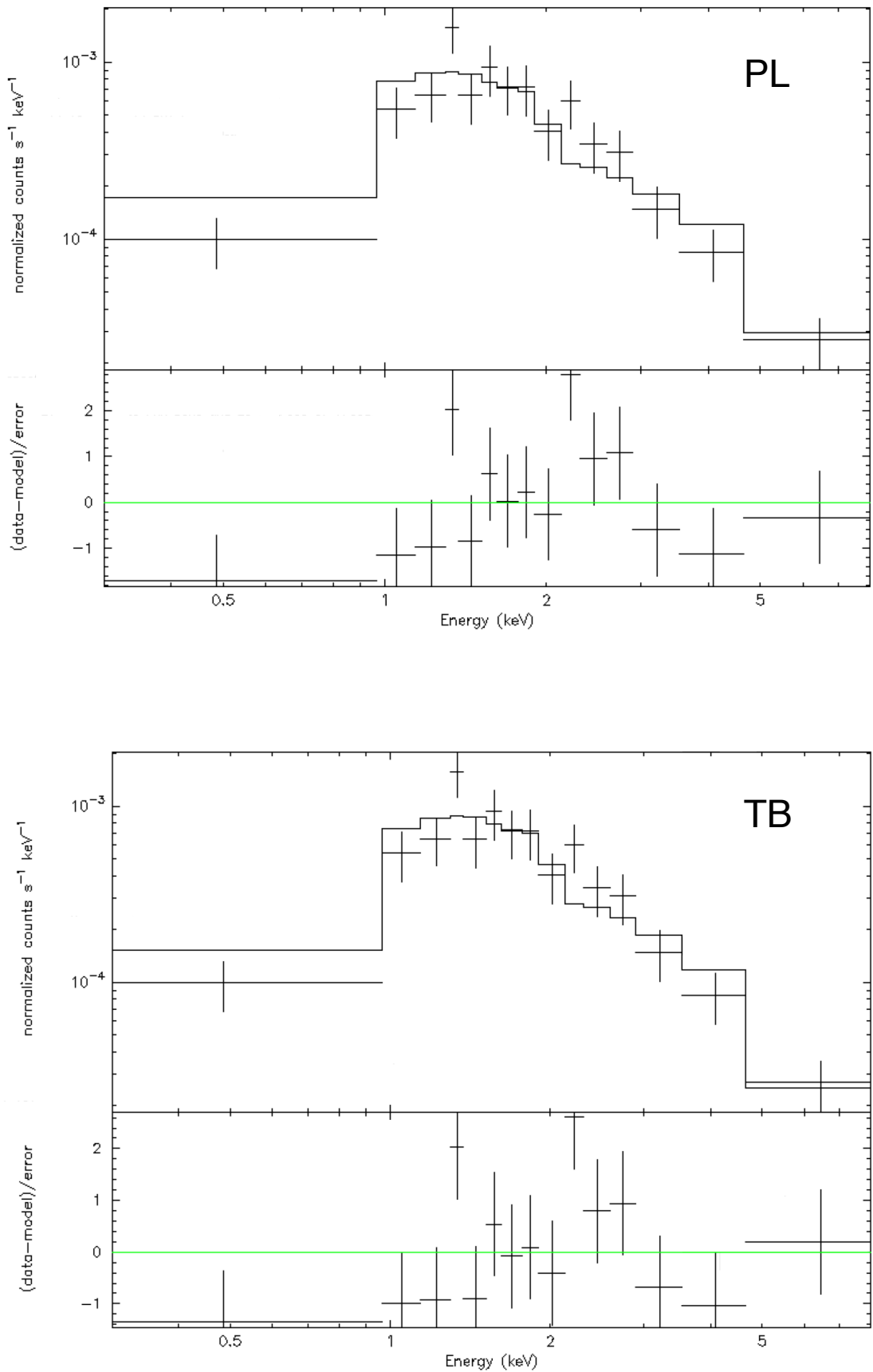

**Fig.2:** The January 2018 merged observations fitted with a power-law (PL, top) and thermal Bremsstrahlung (TB, bottom) model. The column density and redshift were fixed (see text). In each figure, the top panel shows the data with the best fit model. The lower panel shows the deviations from the best fit model. The parameters of both models are summarized in Table 1.

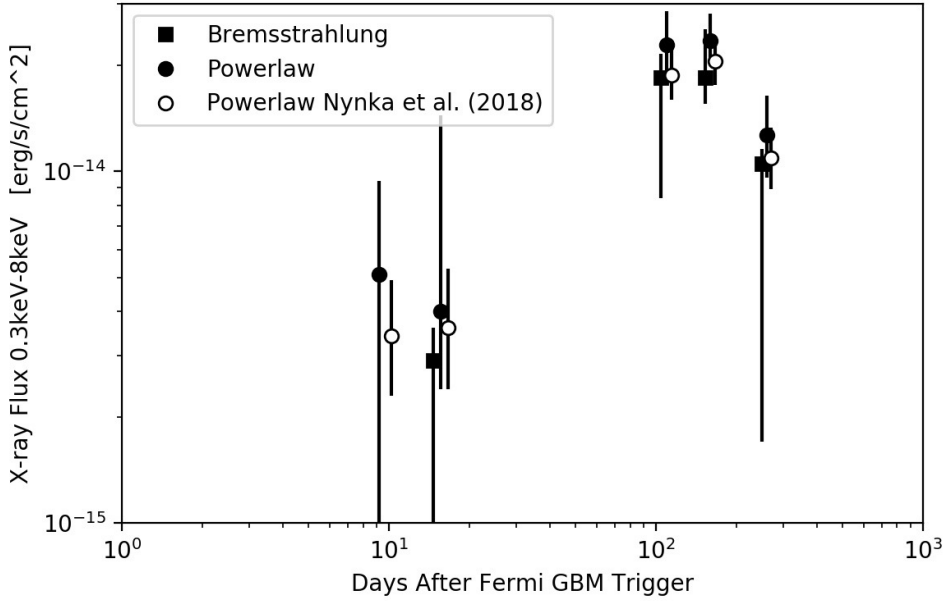

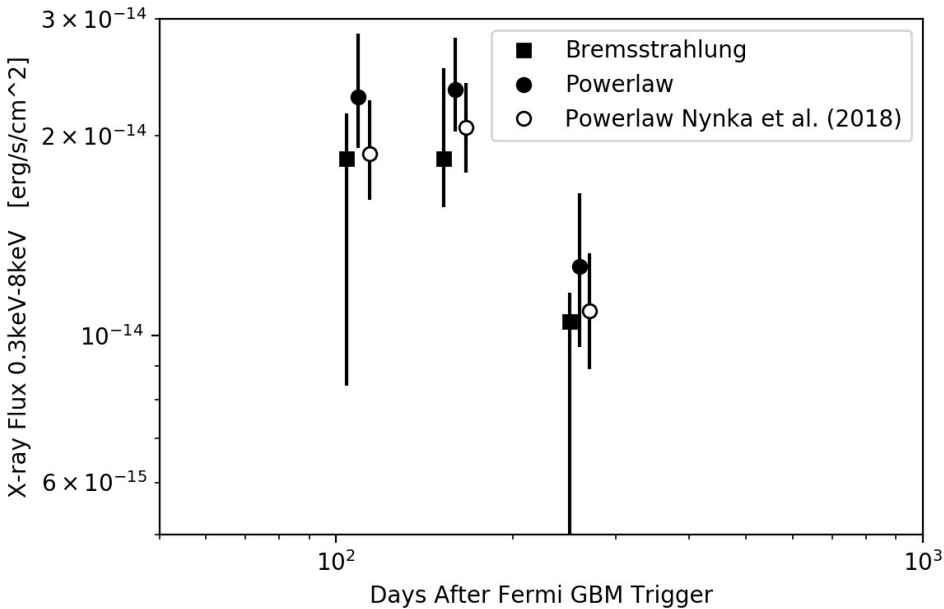

**Fig. 3:** Top shows all five sets of data from Table 1 along with data from Nynka et al. (2018). Below shows the three merged spectra from the latest 3 observation epochs at day >100 post-merger. The open circles are flux values given by Nynka et. al. (2018) using a power-law model. The dark circles are flux values extracted in this work using the same power-law model, and the black square symbols are the flux values extracted in this work using the Bremsstrahlung model. Note that the datapoints are displaced along the time axis only to ease visual inspection.

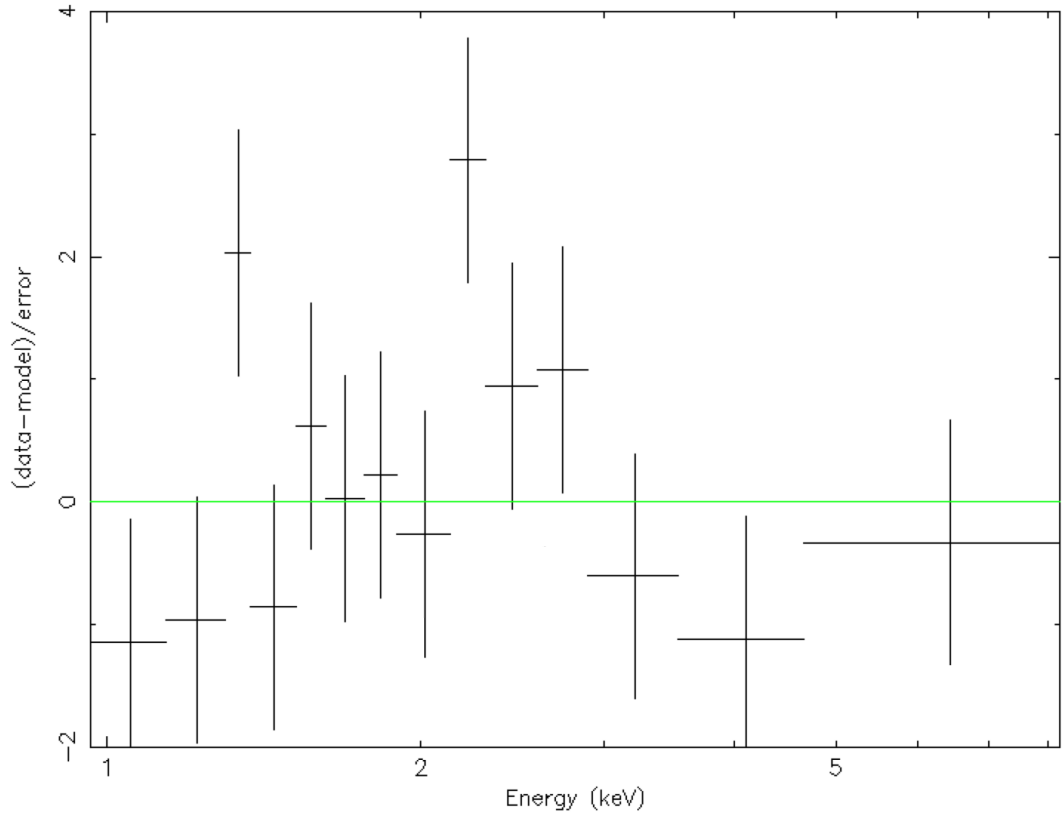

**Fig. 4:** The deviations from an absorbed power law model (green line) for the Jan. 2018 epoch data. The model used is *tbabs*zpowerlw* in XSPEC and the data were grouped by a minimum of 10 counts per bin. The column density was fixed at $N_H=7.5 \times 10^{20}$ cm$^{-2}$ and the redshift at z=0.0098. The data (crosses) show excess near 1.3 keV and 2.2 keV. See the Results section for more details.

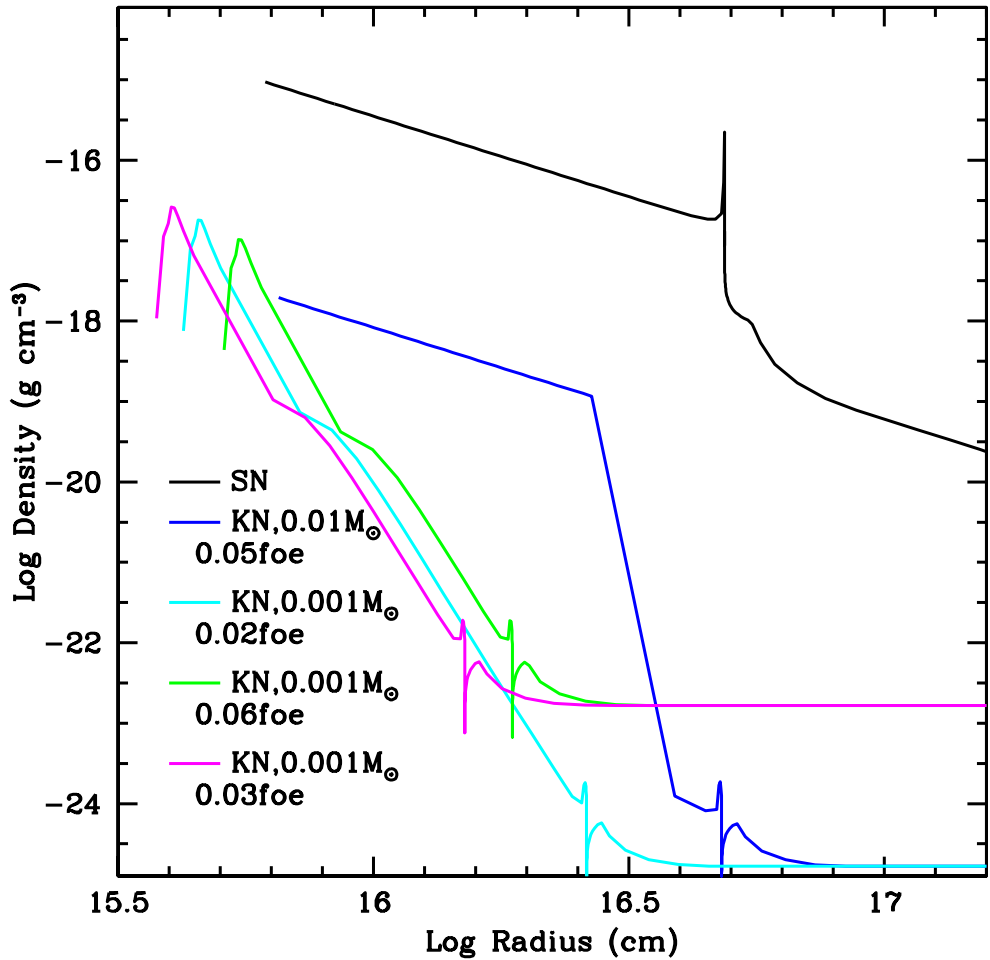

**Fig. 5:** Density versus radius at 100 days for a supernova explosion ($M_{ejecta}$ = 10$M_\odot$, $E_{explosion}$ = 1.5×10$^{51}$ erg, $\dot{M}_{wind}$ = 10$^{-6}$ $M_\odot$ yr$^{-1}$ ) and a suite of kilonova explosions varying the density of the circumstellar medium (0.1, 10 cm$^{-3}$), kilonova ejecta mass (0.001-0.01 $M_\odot$) and kilonova explosion energy (0.02-0.06 x10$^{51}$ erg, or 0.02-0.06 foe). The simulations use a Lagrangian supernova explosion code (Fryer et al. 1996) assuming an ideal gas equation of state. The shock fronts can be seen at positions where the density varies dramatically. At 100 days, the density at the front of the kilonova shock is below 10$^{-22}$ g cm$^{-3}$ where the coupling timescale between ions and electrons becomes long compared to the evolution time of the shock.